\documentstyle[aps,preprint,epsf]{revtex}

\def\lsim{\lower.8ex\hbox{$\buildrel<\over\sim$}}
\def\gsim{\lower.8ex\hbox{$\buildrel>\over\sim$}}
\begin{document}
\draft
\tightenlines

\title{\bf Fokker-Planck Equation for Boltzmann-type and
Active Particles: transfer probability approach}
\author{S. A. Trigger}
\address {Joint Institute for High Temperatures,
Russian Academy of Sciences,\\
13/19 Izhorskaya Str., Moscow 127412, Russia;  \text{email}:strig@gmx.net\\
Humboldt University, 110 Invalidenstr.,
D-10115 Berlin, Germany}
\date{19 December 2002}
     \maketitle

\begin{abstract}
Fokker-Planck equation with the velocity-dependent coefficients is
considered for various isotropic systems on the basis of
probability transition (PT) approach. This method provides the
self-consistent and universal description of friction and
diffusion for Brownian particles. Renormalization of the friction
coefficient is shown to occur for two dimensional (2-D) and three
dimensional (3-D) cases, due to the tensorial character of
diffusion. The specific forms of PT are calculated for the
Boltzmann-type of collisions and for the absorption-type of
collisions (the later are typical for dusty plasmas and some other
systems). Validity of the Einstein's relation for the
Boltzmann-type collisions is analyzed for the velocity-dependent
friction and diffusion coefficients. For the Boltzmann-type
collisions in the region of very high grain velocity as well as it
is always for non-Boltzmann collisions, such as, e.g., absorption
collisions, the Einstein relation is violated, although some other
relations (determined by the structure of PT) can exist. The
generalized friction force is investigated in dusty plasma in the
framework of the PT approach. The relation between this force,
negative collecting friction force and scattering and collecting
drag
forces is established.\\
The concept of probability transition is used to describe motion
of active particles in an ambient medium. On basis of the physical
arguments the PT for a simple model of the active particle is
constructed and the coefficients of the relevant Fokker-Planck
equation are found. The stationary solution of this equation is
typical for the simplest self-organized molecular machines.\\
PACS number(s): 52.27.Lw, 52.20.Hv, 52.25.Fi, 82.70.-y

\end{abstract}
\section{Introduction}
Brownian dynamics nowadays is in the focus of interest due to the
wide new fields of applications: physical-chemical systems,
so-called active walkers, e.g., cells and other objects in
biological systems, dusty plasmas with natural and artificial
grains and many other systems. The characteristic property of such
systems is the velocity dependent friction and diffusion
coefficients. Existence of the Einstein relation and even the
correct specific forms of the Fokker-Planck equation for such
systems are not still completely clarified. In particular, the
attempts to use the Langevin equation as a stochastic basis for
derivation of the Fokker-Planck equation lead to non-sign-valued
result. The different forms of the Fokker-Planck equation, such as
so-called Ito and Stratonovich [1-4] ones, appear. For the systems
close to equilibrium Brownian particles keep stationary random
motion under action of the stochastic forces, which are
compensated by the particle friction and thus, the work produced
by the Langevin sources is equal to the energy dissipated in
course of the Brownian particle motion. This energy balance is
described by the fluctuation-dissipation theorem in the form of
the Einstein law. Obviously, the fluctuation-dissipation theorem
and the Einstein relation can be violated in the case of
non-equilibrium systems (even in the stationary case), in
particular in the open systems. Starting from the classical Lord
Rayleigh work [5] many studies of the non-equilibrium motion of
Brownian particles with an additional (inner, or external) energy
supply have been performed. In particular, such studies are of
great importance for physical-chemical [6,7] and biological [8]
systems, in which non-equilibrium Brownian particle motion is
referred as the motion of active Brownian particles.  The
dynamical and energetic aspects of motion for the active Brownian
particles have been described recently on the basis of the
Langevin equation and the appropriate Fokker-Planck equation
[9,10]. Possibility of negative friction (negative values of the
friction coefficient) for Brownian particles was regarded, as a
result of energy pumping. For some phenomenological dependence of
the friction coefficient as a function of the grain's velocity the
one-particle stationary non-Maxwellian distribution function was
found.

The traditional formulations of the non-equilibrium Brownian
motion are based on some phenomenological expressions for the
friction and diffusion coefficients. In particular, it means that
deviations from the Einstein relation, as well as the velocity
dependence of these coefficients are postulated and high level of
uncertainty for application of such models to the real systems
takes place. Recently we considered another situation, when the
kinetic coefficients can be calculated explicitly on the basis of
microscopically derived Fokker-Planck  equation for dusty plasmas
[11,12].  It was recently shown [13] that in the case of strong
Coulomb interaction of highly charged grains in dusty plasmas, due
to ion absorption by grains, the friction coefficient can become
negative. The necessary criterion for negative friction due to ion
absorption is found as $\Gamma\equiv e^2 Z_g Z_i /aT_i > 1$ (here
$Z_g$, $Z_i$ are the charge numbers for the grains and ions
respectively, $a$ is the grain radius,  $T_i$  is the ion
temperature). The appropriate threshold value of the grain charge
was determined. The stationary solution of the Fokker-Planck
equation with the velocity-dependable kinetic coefficients was
obtained and the considerable deviation of such solution from the
Maxwellian distribution was demonstrated. The physical reason for
manifestation of negative friction in that case is clear:  the
cross-section for ion absorption by grain increases, when the
relative velocity between the ion and grain decreases, due to the
charge-dependent part of the cross-section. Therefore, for a
moving highly-charged grain $(\Gamma\gg1)$ the momentum transfer
from ions to the grain in the direction of grain velocity can be
higher than in the opposite direction.

In this paper we develop more general approach, based on
probability transition, to simplify the Fokker-Planck equation and
to calculate the velocity dependent friction and diffusion
coefficients for the different systems. On that way we find the
various forms of probability transition for the Boltzmann-type and
for absorption collision integrals. The crucial peculiarity of the
exact expressions for the mentioned coefficients follows from the
exact representation of these coefficients through the function of
probability transition:\\
-it is impossible to define the coefficients independently not
only for the processes, which describe the systems close to
thermodynamic equilibrium, when the Einstein relation is
$\grave{a}$ \emph{ priory} valid, but also for the systems in
which there is stationary, but non-equilibrium state,
or for the systems far from equilibrium;\\
-any rigorous approximate model of the Fokker-Planck equation have
to be based on self-consistent expressions for the friction and
diffusion coefficients, based on the PT.

As an example we consider a wide class of open or far from
equilibrium systems, where the Einstein relation is not
applicable. For active particles the suggested consideration can
be easily applied by construction of the probability transition on
basis of the physical arguments.

\section{Probability transition and velocity-dependable friction and
diffusion coefficients}

The appropriate kinetic equation describing motion of Brownian
particles in some medium with the momentum exchange may be written
as
\begin{eqnarray}
\frac{df_g({\bf P},t)}{dt} =  I_g({\bf P},t) = \int d{\bf q}
\left\{w({\bf P+ q, q}) f_g({\bf P+q, t}) \right.\nonumber\\
\left.- w({\bf P, q}) f_g({\bf P},t) \right\},
\label{dust}
\end{eqnarray}
where  $f_g({\bf P})$ is the distribution function of Brownian
particles (grains) of the mass $M$. The elementary process is
change of momentum of the grain ${\bf P}$ to ${\bf (P-q)}$. The
probability transition $w({\bf P, q})$ in Eq.~(\ref{dust})
describes the probability for grain with linear momentum ${\bf P}$
to lose the momentum ${\bf q}$. The Eq.~(\ref{dust}) has a form of
master equation. In general the probability transition is the
function of time itself. To simplify Eq.~(\ref{dust}) for the
processes with the momentum transfer ${\bf q \ll \ \bf P}$ we have
expand the right side of Eq.~(\ref{dust}) by ${\bf q}$. The result
of the expansion is the Fokker-Planck equation for grains:

\begin{eqnarray}
\frac{d f_g({\bf P},t)}{d t} = {\partial \over {\partial P_i}} \left[ A_i({\bf P})
f_g({\bf P}) + {\partial \over {\partial P_j}}
\left( B_{ij}({\bf P}) f_g({\bf P})
\right)\right]
\label{FP}
\end{eqnarray}

The coefficients $A_i({\bf P})$ and $B_{ij}({\bf P})$, as easy to
see by expansion of the Eq.~(\ref{dust}), are expressed
explicitly through the probability transition $w({\bf P, q})$ by
the relations (e.g. [3,14]):

\begin{eqnarray}
A_i ({\bf P})= \int d{\bf q} q_i w({\bf P, q}), \label {A_i} \\
B_{ij}({\bf P})=\frac1{2}\int d{\bf q} q_i q_j w({\bf P,q}).\label
{B_ij}
\end{eqnarray}

Let us suggest that probability transition $w({\bf P, q})$ is a
function of only two vectors ${\bf P}$ and ${\bf q}$. It means,
for example, there is no, let say, drift velocity of the media,
surrounded the grain, as well as some inner vector inside the
grain, which can influence on the probability transition. For that
case the general structure of the coefficients ${A_i}$ and
$B_{ij}$ is evident:

\begin{equation}
\label{scul}
A_i({\bf P})= \ P_i \beta(P),\;\; \\
B_{ij}({\bf P}) = \frac{P_i P_j}{P^2}B_\parallel(P)\ + \,
(\delta_{ij} - \frac{P_i P_j}{P^2})\, B_\perp(P),
\end{equation}
where $\beta (P)$, $B_\parallel(P)\ $ and $B_\perp(P)$ are the
functions of modulus $P$. Let us consider at first the stationary
case to understand the form of the Fokker-Planck equation and
solution, when the friction and diffusion coefficients are the
functions of grain's velocity. On this basis, in particular, the
well known problem related to the Ito [1] and Stratonovich [2]
forms of Langevin and Fokker-Planck equations[3,4] can be solved
and applicability of the Einstein relation for the various kinds
of functions of probability transitions can be investigated. The
results can be used also for unstationary case, if the initial
distribution function is isotropic (what is not always valid,
naturally). Then, taking into account isotropy of the distribution
function $f_g({\bf P,t})$, the Fokker-Planck equation is given by

\begin{equation}
\frac{df_g(V)}{dt} =
{\partial \over {\partial V_i}}
 \left[ \beta^\ast (V) V_i
f_g(V) + \frac{V_i}{V}
{\partial \over {\partial V}}
\left( D_\parallel(V)
f_g(V)\right) \right].\label{eqskal}
\end{equation}
or in the equivalent form:

\begin{equation}
\frac{df_g(V)}{dt} = (s + V
{\partial \over {\partial V}}
) \left[ \beta^\ast (V)
f_g(V) + \frac {1}{V}
{\partial \over {\partial V}}
\left( D_\parallel(V)
f_g(V)\right) \right].\label{eqskal1}
\end{equation}

Here $s$ is the dimension of the velocity space and the scalar
functions of $V$ are the same as ones of $P$, but expressed via
the equality $P=M V$. We use above the velocity variable for
grains ${\bf V}$  instead momentum ${\bf P}$ and the diffusion
tensor $D_{ij}({\bf V}) = M^{-2} B_{ij}({\bf V})$. We also use
these notations below. The functions $\beta^\ast (V)$ and
$D_\parallel(V)$ are determined via the transition probability as

\begin{eqnarray}
\beta^\ast (V) & =& \beta(V)+ \frac{s-1}{V^2}\left
(D_\parallel(V) - D_\perp(V)\right)\, \label{betaa}\\
\beta(V) & =& \frac{1}{P^2}\int d\,^s{\bf q}\, ({\bf Pq})\, w({\bf
P,q}), \label{be} \\
D_\parallel(V) & =& \frac{1}{2M^2 P^2}\int d\,^s{\bf q}\,{\bf
(Pq)}^2\, w({\bf
P, q}), \label{de1} \\
D_\perp(V)& =& \frac{1}{2(s-1)M^2 P^2} \int d\,^s{\bf q}\left[
P^2q^2 - {\bf (Pq)}^2 \right] w({\bf P, q}), \label{de2}
\end{eqnarray}
where ${\bf (Pq)}$ is the scalar product in the velocity space
with dimension $s$. Eq.~(\ref {betaa}) can be rewritten in the
form:
\begin{equation}
\beta^\ast (V) = \beta(V)+ \frac {1}{2P^2} \,\int d{\bf q}\,
w({\bf P, q}) \left[{s \bf (Pq)}^2 /P\,^2 - q^2 \right], \label
{beta^1}
\end{equation}

We see that the three scalar functions of $P$, determined by the
different moments of probability transition, permit to find the
coefficients in the Fokker-Planck equation.

For the anisotropic velocity distribution function or for presence
of the external fields, which do not change the friction and
diffusion (what is not always valid, naturally), Eq. ~ (\ref {FP})
can be rewritten as:

\begin{eqnarray}
\frac{df_g({\bf V)}}{dt} = {\partial \over {\partial V_i}}
 \left[ V_i \beta^\ast (V)f_g({\bf V}) +
 \frac{V_i V_j}{V^2} {\partial \over {\partial V_j}} \left(D_\parallel(V) f_g({\bf V})\right)\right.\nonumber\\
 \left. + \left(\delta_{ij} - \frac{V_i V_j}{V^2}\right){\partial \over {\partial V_j}}\left(D_\perp(V)
 f_g({\bf V})\right)\, \right]. \label{eqskal2}
\end{eqnarray}
For simplicity external fields are not included in Eq.~(\ref
{eqskal2}). The useful equivalent representation of Eq.~(\ref
{eqskal2}) has a form:

\begin{eqnarray}
\frac{df_g({\bf V)}}{dt} = {\partial \over {\partial V_i}}
 \left[ V_i \left(\beta^\ast (V)+ \frac {1}{V}{\partial D_\parallel(V)\over\partial V}\right) f_g({\bf V}) +
 D_\parallel(V) \frac{V_i V_j}{V^2} {\partial f_g({\bf V})\over {\partial V_j}}\right.\nonumber\\
 \left. + D_\perp(V)\left(\delta_{ij} - \frac{V_i V_j}{V^2}\right){\partial f_g({\bf V})\over {\partial V_j}}
\, \right]. \label{eqskal3}
\end{eqnarray}

The stationary solution of the Fokker-Planck equation with the
kinetic coefficients from Eq.~(\ref {betaa},\ref {de1}) for the
grain distribution function $f_g(V)$ is

\begin{equation}
f_g(V) =  \frac{C} {D_\parallel(V)} \exp\left[-\int\limits_0^V
d\upsilon \upsilon \frac
{\beta^\ast(\upsilon)}{D_\parallel(\upsilon)}\right], \label {fgr}
\end{equation}
where $C$ is a constant, providing normalization. As easy to see
from the Eq.~(\ref{eqskal}) for isotropic case the stationary (as
well as non-stationary) Fokker-Planck equation with the
velocity-dependent coefficients has well defined form and the
question of "renormalized" friction coefficient is solved
completely by the Eq.~(\ref{betaa}). The uncertainty in choice of
the Fokker-Planck equation in the forms suggested, e.g., in [1,2]
and [4], created by attempts to connect the Langevin and the
respective Fokker-Planck equations by one-to-one correspondence,
starting from the Langevin equation. The real structure of this
renormalization, due to tensorial character of the diffusion
$D_{ij}({\bf V})$ as follows from the Eq.~(\ref{betaa}), permits
to reformulate the problem: what must be the structure of Langevin
equation for s-dimensional case to be relevant to the
single-valued Fokker-Planck equation, based on the specific
probability transition. Because we use, as the basis, transition
probability we can establish validity or violation of the Einstein
relation between the friction and diffusion coefficients directly,
without usual suggestion of Maxwellian form of the static
distribution function for Brownian particles, which is valid for
the equilibrium state (when the Einstein relation is fulfilled
$\grave{a}$ \emph{priory}). In particular, existence of the
Einstein (or some different from one) relation between the
momentum-dependable coefficients can be investigated.
Correspondence between the Fokker-Planck and Langevin equations
for s-dimensional case on the basis of PT approach will be
considered in detail in a separate publication. Below we find the
probability transition and investigate the various cases for the
PT and Fokker-Planck equations.

\section{The Boltzmann-type collisions}

Let us consider the Boltzmann collision integral for two species
of particles-light component (called below atoms) with the mass
$m$ and grains with the mass $M$ ( $m\,\ll\,M$), which interact
one with another (generalizations can be easily done). To find for
such a process the PT function $w_s({\bf P, q})$ it is enough, for
example, to transform the part of the Boltzmann collision
integral, describing the loss of grains in the phase volume
$d\,{\bf P}$ near the point ${\bf P}$, to the variables ${\bf P}$
and ${\bf q}$, where ${\bf q}$ is the momentum transferred during
the elementary act of collision between atom and grain. Than,
comparing the result of transformation with the Eq.~(\ref{dust}),
for 3D case we find

\begin{eqnarray}
w_s({\bf P,q})= \frac {m}{\mu} \int d{\bf o}\,
\frac{q^2}{2\mu\mid\bf(qn)\mid}\,
\frac {d\,\sigma}{d\,\bf o} \left(\frac{q^2}{2\mu\mid\bf (qn)\mid}, \chi\right)\times \nonumber\\
f_n \left[\frac{m}{M} {\bf P} - \frac{m}{\mu}\left( \frac {q^2 \bf
n}{2 \mid\bf (qn)\mid} - {\bf q}\right)\right], \label {w1}
\end{eqnarray}

Here $d \,{\bf o}= \sin\chi\, d\chi \, d\phi$ is the element of
the space angle for scattering with the differential cross-section
$d\,{\sigma}$, \, $f_n$ is the distribution function of atoms,
${\bf q}= {\bf (p^\prime -p)=(P-P^\prime)}$, \, $q\,\cos\chi= {\bf
(qn)}$, vector ${\bf n}$ is the unit vector along the velocity
${\bf v_0^\prime}$ of atom after collision in the system of center
mass for colliding particles and $\mu$ is the reduced mass. The
values of the limiting angles  for $\chi$ above and below are
usually equal $\chi_{min}=0$ and $\chi_{max}=\pi$, except some
special situations, when the integral over $\chi$ diverges, as for
example, for the purely Coulomb interaction, when the known
cutting with $0<\chi_{min}\leq \chi_{max}< \pi$ is necessary.
Taking into account the relation $ q^2 /\, (2\mid \bf (qn)\mid) =
\mid (\bf v-\bf V)\mid$ Eq.~ (\ref{w1}) can be rewritten in the
form more useful for applications:

\begin{eqnarray}
w_s({\bf V,q}) = \int {d\, \bf \Omega} \int{d\,\bf v} \delta
\left[\; \bf q +\mu\bf v-\mu \mid \bf v \mid \;{\bf n}\;
\right] \times \nonumber\\
 f_n {\bf (v+V)} \,\mid \bf v \mid\;
\frac{d \sigma} {d \,\bf o}\, (\mid \bf v\mid, \chi).\label{ad1}
\end{eqnarray}
We can use the representation (\ref{ad1}) to obtain the useful
general expressions for the friction and diffusion coefficients:

\begin{eqnarray}
\beta(\bf V) = \frac{\mu}{M {\bf V^2}}\int {d\,\bf \Omega}
\int{d\,\bf v} \left[\; \bf V \cdot(\mid \bf v \mid \;{\bf n}-\bf
v)\; \right] \times \nonumber\\
 f_n {\bf (v+V)} \,\mid \bf v \mid
\frac {d\,\sigma}{d\, \bf o}\, (\mid \bf v\mid, \chi).\label{ad2}
\end{eqnarray}
\begin{eqnarray}
D_\parallel(\bf V) = \frac{\mu}{MV^2}\int {d\,\bf \Omega}
\int{d\,\bf v} \left[\; \bf V \cdot(\mid \bf v \mid \;{\bf n}-\bf
v)\; \right]^2 \times \nonumber\\
 f_n {\bf (v+V)} \,\mid \bf v \mid
\frac {d\,\sigma}{d\, {\bf o}}\, (\mid {\bf v}\mid,
\chi).\label{ad3}
\end{eqnarray}
The similar expression can be written for $D_\perp$. When the
distribution function of atoms  $f_n$ has  the Maxwellian form
with the temperature $T$, the function $w({\bf V, q})$ can be
simplified, taking into account the inequality ${\bf
q}$\,$\ll\,{\bf P}$.

If $\mid {\bf V} \mid \;\ll v_T$, where for the thermal velocity
of neutral particles (atoms) we use the notation $v_T$, the
distribution function in PT can be expanded and we arrive, with
respective accuracy of the order $\mu/M\simeq m/M$, at the
expression for $w_s({\bf V, q})$:
\begin{eqnarray}
w_s({\bf V,q}) = -\frac{m}{T}\;n_n \left(\frac{m}{2\pi
T}\right)^{3/2} \int {d\, \bf \Omega} \int{d\,\bf v} \delta
\left[\; \bf q +\mu\bf v-\mu \mid \bf v \mid \;{\bf n}\;
\right] \times \nonumber\\
 exp \left(-\frac{mv^2}{2T}\right) \,\mid \bf v \mid\; (\bf v\cdot \bf V)
\frac{d \sigma} {d \,\bf o}\, (\mid \bf v\mid, \chi).\label{wb}
\end{eqnarray}
If atoms with the density $n_n$ are considered as the point
particles and grains have the radius $a$ it is easy to find from
Eqs.~(\ref {ad2}-\ref{wb}) the values of the coefficients $\beta$
and $D_\parallel$:
\begin{equation}
\beta(V)= 8\frac{\sqrt{2\pi}}{3} (m/M) a^2 n_n v_T,\label{ad4}
\end{equation}
\begin{equation}
D_\parallel(V) = 8\frac{\sqrt{2\pi}}{3} (m/M^2) a^2 n_n T
v_T,\label{ad5}
\end{equation}
We see that the Einstein relation is fulfilled:

\begin{equation}
D(V)= M^{-1}\,T \,\beta (V). \label{E}
\end{equation}
Calculations of $\beta$ and $D_\parallel$ were done independently
on the basis of the appropriate PT function.

Let us now consider under general condition  ${\bf q}$\,$\ll\,{\bf
P}$ the opposite case $\mid {\bf V} \mid \gg v_T$ to solve the
problem of validity of the Einstein relation for the velocity-
dependable coefficients of the Fokker-Planck equation for
arbitrary values of grain velocities [4]. In fact the answer can
be found already for the particular case of atoms, scattering by
grains, considering as the hard spheres. Even in this simple case,
as it will be shown, the Einstein relation is violated for high
grain velocity. In the limit of extremely high grain velocity we
can use the simplest approximation for the distribution function
$f_n ({\bf v} +{\bf V})= n_n \delta (\bf v+\bf V)$ and from Eqs.
~(\ref{de2},\ref{ad1}-\ref{ad3}) obtain:

\begin{equation}
\beta(V)= (2m /3M) \pi a^2 n_n V, \label{ad6}
\end{equation}
\begin{equation}
D_\parallel(V) =  (m /M^2) \pi a^2 n_n V^3,\label{ad7}
\end{equation}
For $D_\perp$ by similar calculations we find
\begin{equation}
D_\perp(V)= \frac {D_\parallel(V)}{4} \; , \label{ad8}
\end{equation}
Therefore in that limit instead the Einstein relation we find
another relation between velocity-dependable $\beta(V)$ and
$D_\parallel(V)$:
\begin{equation}
D(V)= \frac{2}{3M} m V^2 \,\beta (V). \label{ad9}
\end{equation}
This situation can be classified as far from equilibrium. For the
considering case of uncharged spherical grains the simplest
interpolation relation between the friction and diffusion
coefficients can be suggested:
\begin{equation}
D(V)= \frac{T}{M} \left(1 + \frac{2m V^2}{3T}\right) \,\beta (V).
\label{ad10}
\end{equation}
The similar interpolations as well as the exact velocity-dependent
relations for the arbitrary cross-sections can be found from Eqs.~
(\ref{ad2}), (\ref{ad3}). As it follows from the relations
considered above $\beta^\ast (V)\simeq\beta(V)$ for all the cases
with accuracy $m/M$.  In general, by use the expansions to higher
order of the small relation $q/P$,  the expressions for the
functions $\beta^\ast (V)$, $D_\parallel(V)$ and $D_\perp(V)$ can
be calculated and possibility to neglect the difference between
$\beta^\ast (V)$ and $\beta (V)$ can be established for the
considered Boltzmann-type collisions.

The essential influence of velocity dependence on the values of
the friction and diffusion coefficients and violation of the
Einstein relation for the Boltzmann type collisions takes place
only for the extreme velocities much higher than the thermal
velocity of the light particles. For the case of low grain
velocity on basis of general representation (\ref{ad2}),
(\ref{ad3}) for the friction and diffusion coefficients it can be
shown applicability of the Einstein relation for arbitrary
cross-section of scattering.

\section{Absorption collisions}

Now turn to other type of collisions, namely to the absorption
collisions, which are typical, for example, for dusty plasmas and
some other open systems. As well known the process of grain
charging by absorption of the electrons and ions by grains leads
to the stationary (but non-equilibrium) state in plasma
discharges. In so called OML (orbital motion limited)
approximation the electrons and ions approaching to the grain on
the distance less than the grain radius $a$ are assumed to be
absorbed. It is clear that the absorption collisions cannot be
described by the Boltzmann-type collision integral. The
appropriate correct form of the absorption collision integral have
been postulated and applied in [14,15]. The rigorous kinetic
theory of the electron and ion absorption in dusty plasmas, which
exists in parallel with the usual processes of electron and ion
scattering by grains, have been developed in [11,12], where also
the Fokker-Planck equation for the charged grains was justified.
Below we use the simplest form of the Fokker-Planck equation for
grains with a fixed charge (distribution by charge assumed narrow,
what is often in reality). Our aim here is to find the probability
transition function for absorption and to demonstrate efficacy of
such approach. More complicated cases can be considered similarly.
We also ignore increase of the grain mass [16, 17], assuming that
neutral atoms generated in the course of the surface electron-ion
recombination escape from the grain surface into a plasma.
Naturally this process also changes the momentum balance for the
particles, but we will not consider this process in our present
model. Then the kinetic equation for grains can be written as:

\begin{eqnarray}
\frac {df_g({\bf P},Q,t)}{dt} =  I_g({\bf P},Q,t)= \hfill
\nonumber\\
\sum_\alpha \int d{\bf p} f_\alpha({\bf p}) \left\{W_\alpha({\bf
p} , {\bf P- p} , Q) f_g({\bf P-p},Q) \right. && \left. -\;\;
W_\alpha({\bf p} , {\bf P} , Q)
 f_g({\bf P},Q) \right\},
\label{pldust}
\end{eqnarray}
where $f_\alpha({\bf p})$ is the distribution functions for the
electrons ($\alpha=e$) and ions ($\alpha=i$) and $Q=eZ_g$.  The
elementary process is the absorption of electron or ion with the
mass $m_\alpha$. The probability of absorption is given by:

\begin{equation}
W_\alpha({\bf p} , {\bf P} , Q)=\sigma_c \left(\left|\frac{\bf
P}{M} -\frac{\bf p}{m}\right| , Q \right) \left| \frac{\bf P}{M}
-\frac{\bf p}{m_\alpha} \right|, \label{w3}
\end{equation}
where $\sigma_c(v, Q)$ with $v \equiv(\mid\frac{\bf P}{M}
-\frac{\bf p}{m}\mid)$ is the cross-sections for absorption (or
collection) of the light plasma particles by grain in OML theory:

\begin{equation}
\sigma_c(Q,v) = \pi a^2 \left(1-{2e_\alpha Q\over m_\alpha v^2a}
\right) \theta \left(1-{2e_\alpha Q\over m_\alpha v^2 a}\right).
\label{si}
\end{equation}

Below on the basis of the Eqs.~(\ref{pldust})-(\ref{si}) two
different, but related problems are solved. At first, to consider
the problem of absorption in simplest form, we accept the next
simplification in spirit of [16,17], namely we will consider
absorption of neutral atoms of a one sort. It means we put $Q=0$
and instead summation by $\alpha$ save only notation with the
index $n$ (for the neutral atoms) in the
Eqs.~(\ref{pldust})-(\ref{si}). Generalization of this simplest
model to the case of charge absorption with a fixed charge is
quite simple.

The second problem in focus of our interest is connected with a
real dusty plasma, when there is ion stream and so called drag
force, applied to the grains and created by ion absorption and ion
scattering, exists.

Let us start with a system of neutral particles, absorbing by
grains. The momentum transferred to the grain due to absorption,
is equal to the momentum of the atom ${\bf p}$ colliding with the
grain. Therefore the probability transition $w_c({\bf P, q})$ for
the considered case can be immediately found by comparison of the
Eqs.~(\ref {dust}), (\ref{pldust})

\begin{equation}
w_c({\bf P, q})= f_n({-\bf q})\,\sigma_c \left(\left|\frac{\bf
P}{M} + \frac{\bf q}{m}\right| \right)  \left| \frac{\bf P}{M}
+\frac{\bf q}{m} \right|, \label{w4}
\end{equation}

If we choose the Maxwellian distribution for atoms and suppose, to
consider the simplest case, that absorption is purely geometrical
$\sigma_c = \pi a^2$ \,\,(the particular case of the
Eq.~(\ref{si}) for $Q=0$), \,we easily obtain with accuracy $\sim
m/M$, that $\beta^\ast (V)\simeq\beta(V)$. A simple calculation
leads to the relation between $\beta(V)$ and $D_\parallel(V)$:

\begin{equation}
D_\parallel(V)= 2M^{-1}T \beta(V) \label{E1}
\end{equation}

This relation is different from the Einstein one already for low
$V$ and for $V=0$ coincides with the result obtained in [12,17],
as the limiting case ($Q=0$) of the Fokker-Planck equation for
dusty plasma in the case of the dominant absorption collisions.
Here we found this relation, on basis of the theory with
velocity-dependent coefficients, based on the probability
transition approach, developed above. For more general form of the
probability transition, which contains the free functions
$\psi(P)$,$\chi(q)$ and a small parameter $\zeta\ll\,1$, $w({\bf
P, q})= \psi (P)\,\phi (\left|{\bf q+\zeta P}\right|)\,\chi(q)$ it
is possible to show, that as above (with accuracy to $\sim \,m/M$)
$\beta^\ast (V)\simeq\beta(V)$ and the relation between $\beta(V)$
and $D_\parallel(V)$ also has a form independent from $V$ and
different from the Einstein one

\begin{equation}
\frac {\beta(V)}{D_\parallel(V)}= \frac {2\zeta \, M^2\int d{\bf
q}\,q\chi(q)\, (\frac{\partial\phi(q)} {\partial q})} {\int d{\bf
q}\,q^2\chi(q)\,\phi(q)}. \label{E2}
\end{equation}

This consideration shows that the structure of the Fokker-Planck
equation for the processes, based on the Boltzmann-type collision
integrals, is very different from the processes of other type,
when the Boltzmann-type collisions are not relevant. For the first
type of processes the Einstein relation is valid, even for the
case of velocity dependent the friction and diffusion
coefficients, but in the limit of low grain velocity $\mid {\bf
V}\mid \ll v_T$ . For non-Boltzmann type of momentum transferring
the fluctuation-dissipation theorem does not exist even for $V=0$,
through some relation between the friction and diffusion
coefficients can take place (specific for the each type of PT).
These results have deep consequences for many physical systems, as
well as for the systems of biological nature, e.g., cells, moving
in solutions and other, so called, active walkers.

\section{Friction and drag force in dusty plasma}

In this section as an application of the developed theory we
consider the problem of momentum transfer from the ion stream to
grains in dusty plasmas. Due to ion absorption and scattering by
grains the drag force, acting on grains, appears. This force plays
a crucial role in many experimental observations and, probably, is
important for formation of voids in dusty plasmas for both
"earthly" [18-20] and microgravity [21] conditions; the theory of
voids essentially based on the drag force was given in [22,23].
The ion drag force ${\bf D}_{I}$ consists of two parts, so called
collection ${\bf D}_{Ic}$ and scattering ${\bf D}_{Is}$ ones. In
the paper of Barnes et al.[24] the approximate analytical
expressions for the drag collection and scattering forces were
done. Later on the theory of drag force was actively developed
analytically and numerically [23-28] to improve the description of
drag force. It was achieved for the scattering part ${\bf
D}_{Is}$, in particular, in the recent publications [27,28].

Here we are focusing on generalized description of friction and
relation between the friction and drag forces. Let us calculate
the generalized expression for the friction force ${\bf F}_f$ on
basis of the Fokker-Planck equation for the charged grains
((Eqs.~(\ref{A_i}),(\ref{pldust})). We calculate here ${\bf F}_f$
only for ions, because we are interested the ion part of friction.
Generalization for many species of the light components
(electrons, atoms) is simple. By integration of Eq.~(\ref{FP}) on
momentum we find for the time evolution of the average momentum
${\bf P}_g$:

\begin{equation}
\frac {\partial\,(n_g {\bf P}_{gi})}{\partial\, t} =
 -\,\int d\,{\bf P}\,\tilde {A}_i ({\bf P, y})\,f_g({\bf P},\,t), \label{FRIC}
\end{equation}

The function $\tilde {A}_i ({\bf P, y})$ is a generalization of
${A}_i ({\bf P})$ of the Eq. (\ref{scul}) for the case of
existence of some additional vector in the probability transition.
In the case under consideration this vector ${\bf y}$ is the
momentum, determined by the velocity of the ion stream ${\bf y}= M
{\bf u}$:

\begin{equation}
\tilde {A}_i ({\bf P, y})=   \int d{\bf q}\, q_i\, \tilde w({\bf
P,y, q}), \label {A_u}
\end{equation}

To find $\tilde w({\bf P,y, q})$, when there is ion flow in
plasma, it is quite natural to use the shifted Maxwellian
distribution function of ions.

At first we consider the scattering part ${\bf F}_{fs}$ of ${\bf
F}_f$. The probability transition $\tilde w_s({\bf P,y, q})$ in
this case has the evident property:
\begin{equation}
\tilde w_s({\bf P,y, q})= w_s({\bf P-y, q})\label {P1}.
\end{equation}
Then we obtain the general expression for the friction force ${\bf
F}_{fs}$:

\begin{eqnarray}
{\bf F}_{fs} &=& - \int d{\bf P}\,
{\bf(P-y)}\, \beta (\left|{\bf P-y}\right|)\, f_g(P,t), \label {Ff}\\
\beta {\bf (P-y)}&=& \frac {1}{({\bf P-y})^2}\, \int d{\bf
q}\,{\bf [\,q\,\cdot(P-y)\,]}\;w_s({\bf P-y, q}), \label {Ff1}\\
{\bf \tilde A {\bf (P-y)}}&=& ({\bf P-y}) \beta (\left|{\bf
P-y}\right|)= \int d{\bf q}\, {\bf q}\, w_s({\bf P-y, q}), \label
{AP}
\end{eqnarray}
In the limit cases of the scattering friction itself ${\bf
F}_{f0s}\equiv {\bf F}_{fs}^>$ and scattering ion drag itself
${\bf D}_{Is}\equiv{\bf F}_{fs}^<$ the Eq.~(\ref{Ff}) can be
rewritten as:

\begin{equation}
{\bf F}_{fs}^>= - \int d{\bf P}\, {\bf P}\, \beta (\left|{\bf
P}\right|)\, f_g(P,t), \qquad {\bf P}\gg {\bf y} \label {Ff2}
\end{equation}
\begin{equation}
{\bf F}_{fs}^< = n_g {\bf y}\, \beta (\left|{\bf y}\right|)\, ,
\qquad {\bf P}\ll{\bf y} \label {Ff3}
\end{equation}
For the momentum distribution function of grains $f_g = \delta
({\bf P-P_0})$ (if the force is calculated for one particle)
Eq.~(\ref{Ff}) takes a form:

\begin{equation}
{\bf F}_{fs}= - ({\bf P_0-\bf y})\, \beta (\left|{\bf P_0-\bf
y}\right|), \label {Ff4}
\end{equation}
and describes both the friction force itself and the drag force.
In general, according to Eq.~(\ref{Ff}), (\ref{Ff4}), there is
competition between friction and acceleration. For the limiting
case of the friction force itself ${\bf F}_{f0s}$ for the grain
with momentum $\bf G$ and immobile ions and the opposite case -
ion drag itself ${\bf D}_{\emph{I\,s}}$ with the ion velocity $\bf
u= \bf G /M$ there is natural relation:
\begin{equation}
{\bf F}_{f0s}\equiv {\bf F}_{fs}^>(\bf G)= -{\bf
F}_{\emph{f\,s}}\,^< (\bf G)\equiv {\bf D}_{\emph{I\,s}}\,(\bf G),
\label {Ff5}
\end{equation}
This picture can be easily generalized for a few species of the
light particles. From Eqs.~(\ref{ad1}), (\ref{Ff4}) we find for
${\bf F}_{fs}$ the representation:
\begin{equation}
{\bf F}_{fs} =  \frac {m \bf (V-u)}{(\bf V-\bf u)^2} \int d{\bf v
}\, f_i {\bf (v-V+u)}[{\bf(u-V)}\cdot{\bf v}]\left|{\bf
v}\right|\,\sigma _{tr} ({\bf v}) , \label{Fs}
\end{equation}
\begin{equation}
\sigma _{tr} ({\bf v}) = \int_{\chi_{min}}^{\chi_{max}} d\,{\bf
\Omega}\, \frac {d\,\sigma}{d\,\bf \Omega} (1-\cos\chi), \label
{Ff6}
\end{equation}
Here we use the limits for the  angles of integration, taking into
account to provide convergence for the Coulomb cross-section in
the case of ion-grain scattering. Eq.~(\ref{Fs}) coincides in the
limit case $\bf V=0$ with the well known general formulae for the
transferring of momentum from light to heavy particle in the
process of scattering, which can be justified from the simple
physical arguments, as it was done, e.g., in [29]. For the
opposite limit case $\bf u=0$ it describes the friction force
${\bf F}_{f0s}$ for grain. This equation is also applicable,
naturally, for the short-range scattering potentials, when
$\chi_{min}=0$ and $\chi_{max}=\pi$. The specific result for ${\bf
D}_{\emph{I\,s}}$ can be written for the Coulomb cross-section in
the form:
\begin{eqnarray}
{\bf D}_{\emph{I\,s}}= 2A_0 M {\bf u}\Gamma^2 \ln\Lambda .
\label{ad11}
\end{eqnarray}
where $A_0=\frac{\sqrt{2\pi}}{3} (m_i/M) a^2 n_i v_{T_i}$ and
$v_{T_i}\equiv\sqrt{T_i/m_i}$. The parameter $\Gamma \equiv e^2
Z_g Z_i /a T_i$ and usually $\Gamma\gg1$ for dusty plasmas. The
structure of the generalized Landau logarithm $\ln\Lambda$ for
dusty plasma is very important and have been recently considered
in [27,28]. For very strong interaction the problem of the correct
form of $\ln\Lambda$ as function of plasma parameters is still not
completely solved.

Let us consider now the generalized collecting friction ${\bf
F}_{fc}$ and in particular the collecting drag force ${\bf
D}_{Ic}$. The formula of the similar to Eq.~ (\ref{Fs})structure,
but with ${\bf V}=0$ and with the collecting non-transport
cross-section, instead transport scattering cross-section, was
applied also for the collecting drag force in [23,27,28] and other
papers on phenomenological basis. Our goal here to investigate the
collecting drag force by use the PT Eq.~(\ref{w4}) for absorption
and to find the relation between the friction force and the
friction coefficient for the collecting process. Recently in [13]
the friction coefficient $\beta(V)$ in dusty plasmas was
calculated explicitly for arbitrary grain velocity and parameter
$\Gamma $. It was found, that $\beta(V)$ can change sign
("negative friction") from positive to negative for some velocity
domain if the parameter $\Gamma > 1$. Here we reproduce the result
of [13] for the total friction coefficient for a grain $\beta(V)$
(for the particular but important case $\eta\equiv(m_i
v^2/2T_i)\ll1$ and arbitrary $\Gamma$):

\begin{eqnarray}
\beta(\eta,\Gamma) &=& 2A_0 \left[1-\Gamma+ 4{n_a\over n_i}\,
\left({T_a m_a\over T_i m_i}\right)^{1/2}
\right.\nonumber \\
&-& \left. {\eta\over5} (1-3\Gamma) + \Gamma^2\ln\Lambda
\right].\label{Ad1}
\end{eqnarray}
The terms in Eq.~(\ref{Ad1}) are proportional to the atom density
$n_a$ and to $\ln\Lambda$ describe friction respectively with
atoms and with ions by scattering. These terms are always
positive. Other terms in Eq.~(\ref{Ad1}) describe "negative
friction" due to ion absorption by grains and are negative in the
considering limit case $\eta\ll1$ if $\Gamma > 1$. "Negative
friction" exists for small $\eta$, if the Coulomb scattering is
strongly suppressed, when the Coulomb logarithm $\ln\Lambda$ is
small [13,27] (some additional reasons for it reduction are
discussed in [28]), that is typical for strong interaction in
dusty plasmas. The level of ionization has to be enough high to
provide negative value of the friction coefficient. As we know
these conditions in present are not reached in the experimental
set-ups. Opportunity for manifestation of "negative friction" in
the experiments requires as we already mentioned the special
conditions. From Eqs.~(\ref{FRIC}),(\ref{A_u}) and (\ref{Ad1}) it
follows straightforward that the collection friction force itself
${\bf F}_{f0c}$ for a moving grain can be written as:
\begin{equation}
{\bf F}_{f0c} = - {\bf P_0}\beta(\eta,\Gamma) .\label{ad12}
\end{equation}

To find the generalized collecting friction force ${\bf F}_{fc}$
(and therefore also the drag force) we have find the function
$\tilde w_c({\bf P,y, q})$ for collection. The crucial fact is
that the relation similar to Eqs.~(\ref{P1}) or (\ref{ad12}) for
PT function is not correct for the absorption in the case when an
ion flow exists ($\bf u\neq0$), due to the different structure of
the PT functions for the scattering and collection processes.

To describe the ion stream with the velocity ${\bf u}\neq0$ we use
again the shifted Maxwellian distribution of ions. As it easy to
see in this case the PT function $\tilde w_c({\bf P,y, q})$ can be
expressed via $w_c$ determined by Eq.~(\ref{w4}):
\begin{equation}
\tilde w_c({\bf P,y,q}) = w_c ({\bf P-y,q + \emph{m}
u}).\label{ad13}
\end{equation}
For the momentum distribution function of grains $f_g = \delta
({\bf P-P_0})$ (if the force is as above calculated for one
particle) the friction ${\bf F}_{fc}$ takes the form:
\begin{equation}
{\bf F}_{fc} = - \tilde {\bf A}(\bf P_0, y)= -\int dq \,({\bf q}-
\emph{m }{\bf u})\, \ \emph{w}_c ({\bf P_0-y},\,{\bf q})
.\label{ad14}
\end{equation}
This equality can be written in the equivalent form:
\begin{equation}
{\bf F}_{fc} = m{\bf u}\beta_1(\mid {\bf P_0-y}\mid)- ({\bf
P_0}-m{\bf u})\beta_2(\mid{\bf P_0-y}\mid),\label{ad15}
\end{equation}
where the coefficients $\beta_i$ are related with the zero and
first moments of the PT function:
\begin{equation}
\beta_1 = \int d{\bf q} \,w_c ({\bf P_0-y},{\bf q}) \label{ad16}
\end{equation}
\begin{equation}
\beta_2 = \frac {1}{{\bf (P_0-y)}^2}\int d{\bf q}\, [{\bf
q}\,\cdot{\bf (P_0-y)}]\,w_c ({\bf P_0-y},{\bf q}) \label{ad17}
\end{equation}
Let us consider the simple and practically important case, when
the both vectors ${\bf P_0}$ and ${\bf y}$ are directed parallel
or antiparallel to the same unit vector $\bf l$: ${\bf P_0}=
{\pi}_0 {\bf l}$  and ${\bf y}= y_0 {\bf l}$. Then for the
friction we arrive at the following expression:
\begin{eqnarray}
{\bf F}_{fc} = {\bf l}\left\{u_0 \left[ m\beta_1
({\mid\pi}_0-y_0\mid)+M \beta_2({\mid\pi}_0-y_0\mid)\right]-
{\pi}_0 \beta_2(\mid{\pi}_0-y_0\mid) \right\}, \label{ad18}
\end{eqnarray}
Here and below we use velocities related with the momenta ${\pi}_0
\equiv M V_0$ and ${y_0} \equiv M {u_0}$. Eq.~(\ref{ad18}) can be
represented in the equivalent and explicit form:
\begin{eqnarray}
{\bf F}_{fc} = - {\bf l}\left\{m u_0 \int {\bf dv}\, \frac{{\bf
v}\cdot {\bf l}\, (u_0-V_0)}{(u_0-V_0)^2} f_i [{\bf v}+{\bf
l}(u_0-V_0)]v \sigma_c
(v)  \right.\nonumber\\
\left. - m V_0 \int {\bf d v}\, \frac{[{\bf v} \cdot {\bf l}+
u_0-V_0]\, (u_0-V_0)}{(u_0-V_0)^2} f_i [{\bf v}+{\bf l}(u_0-V_0)]v
\sigma_c (v)  \right \}, \label{ad19}
\end{eqnarray}
Let us consider the special cases of Eq.~(\ref{ad19}). \\
a) $V_0 \gg u_0, u_0\rightarrow0$. In this case we arrive to the
friction force ${\bf F}_{f0c}$:
\begin{eqnarray}
{\bf F}_{f0c} = - m {\bf V}_0\int d {\bf v} \frac{[{\bf V}_0 \cdot
({\bf v}-{\bf V}_0)]}{V_0^2} f_i [{\bf v}-{\bf V}_0]\,v \sigma_c
(v), \label{ad20}
\end{eqnarray}
This expression coincides with the collecting ion friction force,
which leads to the negative collecting friction coefficient for
$\Gamma>1$ [13] and to the respective relations (\ref{Ad1}),
(\ref{ad12}).

b) $u_0 \gg V_0, V_0\rightarrow0$. In this case the generalized
collecting friction ${\bf F}_{fc}$ describes the collecting ion
drag force ${\bf D}_{\emph{I\,c}}$\,:
\begin{eqnarray}
{\bf D}_{\emph{I\,c}}=  m {\bf u}\int d {\bf v} \frac{{\bf u}
\cdot {\bf v}}{u^2} f_i [{\bf v}-{\bf u}]\,v
\sigma_c(v)\label{ad21}
\end{eqnarray}
This equation coincides with the expression for the collecting
drag force, which have been suggested in [12] and used also in
[27,28].

c) $u_0\neq0, V_0\neq0$. Temperature of ions is low
$\mid(u_0-V_0)\mid \gg v_{Ti}$, the ion distribution function
tends to delta function $n_i \delta({\bf v}+{\bf u}_0-{\bf V}_0)$.
Then ${\bf F}_{fc}$ tends to the force directed along the ion
stream, which we denote as ${\bf D}_{I0c}$:
\begin{eqnarray}
{\bf D}_{I0c}= m n_i {\bf u} \mid(u_0-V_0)\mid
\sigma_c(\mid(u_0-V_0)\mid), \label{ad22}
\end{eqnarray}
As it follows from Eq.~(\ref{ad22}) ion wind in the considered
limit always accelerates grains. If $u_0$ is parallel to $V_0$ and
they are close one to another, but both are higher than $v_{Ti}$,
the enhancement of the drag force ${\bf D}_{I0c}$ occurs with
decrease of the relative velocity $\mid(u_0-V_0)\mid$. It is the
consequence of the OML collecting cross-section, which probably
can be observable for very fast grains, discovered in some
experiments [30,31], or in cryogenic discharges.

\section{Active particles}

During the last decade investigation of motion of the
self-organized objects, as, e.g., cells, is in a focus of
interest, due to numerous  measurements and observations of their
dynamical behavior [8,32]. Our goal in this work to show, that
construction of the relevant probability transition on basis of
the simple physical requirements permits to justify the relevant
description of such systems. In particular, we show, that for the
simplest structure of PT for motion of an active particle, the
known (and experimentally verified) structure of the velocity
distribution of grains (cells) [32], which are able to the
directed motion near a fixed non-zero velocity, can be justified.
The coefficients of this distribution are calculated.

Let us formulate some general conditions to find the structure of
the PT for active particles. We suppose that the linear momentum,
transferred from a grain (cell) to the surrounding medium, is
created by loss of the inner energy of this grain. Below we ignore
the processes of energy supply, which can be included separately
in more complicated schemes.

At first we assume that the transferred to medium momentum ${\bf
q}\ll{\bf P}$ is distributed near some fixed value ${\bf q_0}$.
The frequency of generation of the transferring momentum {\bf q}
will be denoted as $\mu({\bf q})$. It can be approximated, for
example, by the product of the functions, describing distributions
on modulus $\mid{\bf q}\mid=q$ and on the space angles
$\theta,\varphi$ between the vectors ${\bf q}$ and ${\bf P}$, if
there are no others, besides ${\bf q}$ and ${\bf P}$,
characteristic vectors for the system. In this simplest case we
can put $\mu({\bf q})=\nu_0\,\mu_1(q)\mu_2(\theta,\varphi)$, where
$\nu_0$ is the q-independent frequency of momentum generation by a
grain (cell). Below we suggest that the $\varphi$-dependence of
$\mu_2$ is absent. The distribution $\mu_1(q)$ can be gaussian or,
for the limit case of very narrow distribution of the transferring
momentum, can be approximated by the delta-function $\mu_1(q)=
\lambda_0\delta(q-q_0)/\ q_0^2$, where $\lambda_0$ is a
dimensionless constant. For the function $\mu_2(\theta)$ we
suppose, that the angle $\theta$ between the direction of the
transferred momentum and the momentum of the active grain is
enclosed between the values $\pi-\theta_0<\theta<\pi$, where
$\theta_0$ is some acute angle. Due to this amplification of the
grains takes place. Consideration for the 2D and 1D cases is
evident.

The "weight function" $\Delta(\theta)$ for the angle $\theta$ can
be included to describe the axis-symmetrical non-homogeneity of
amplification for the different angles $\theta$. We can also
include two additional weight "functions": $\Sigma(P)$ and
$\Upsilon(\varepsilon-\varepsilon_0)$. The first one describes
dependence PT from the modulus $P$ of the momentum ${\bf P}$, the
second one provides that momentum transferring is possible only if
the inner energy of a grain (cell) $\varepsilon$ is bigger than
some fixed minimal value of the inner energy $\varepsilon_0$, let
say $\Upsilon(\varepsilon-\varepsilon_0)=\vartheta
(\epsilon-\varepsilon_0)$, where $\vartheta$ is the step-like
function.

Under these assumptions the PT for the momentum transferring
$w_\varepsilon$, due to loss of the inner energy of an active
particle, can be written in the case under consideration as:

\begin{equation}
w_\varepsilon({\bf P, q})=
\nu_0\Upsilon(\varepsilon-\varepsilon_0)\Sigma(P)\Delta(\theta)\mu_1(q)
\vartheta(\theta-\pi+\theta_0) \vartheta(\pi-\theta).\label{AC1}
\end{equation}
The values of the friction and diffusion coefficients follows from
general Eqs. (\ref{betaa})-(\ref{de2}):

\begin{eqnarray}
\beta_\varepsilon (V) &=& -2\pi\nu_0
\Upsilon(\varepsilon-\varepsilon_0)\frac{\Sigma(P)}{P}
\int\limits_0^\infty d q q^3 \mu_1(q)
\int\limits_{\pi-\theta_0}^\pi d\theta\sin\theta\cos\theta\Delta(\theta), \label{AC3}\\
D_\varepsilon(V)&=& 2\pi\nu_0
\Upsilon(\varepsilon-\varepsilon_0)\frac{\Sigma(P)}{M^2}
\int\limits_0^\infty d q q^4
\mu_1(q)\int\limits_{\pi-\theta_0}^\pi
d\theta\sin\theta\cos^2\theta\Delta(\theta) \label{AC4}
\end{eqnarray}
To find the total expressions for $\beta (P)$ and $D(p)$ we have
add to the values (\ref{AC3}),(\ref{AC4}) of $\beta_\varepsilon
(P)$ and $D_\varepsilon(P)$ the parts of the friction and
diffusion coefficients aroused due to collisions between the cell,
moving with the momentum P, and the surrounding particles (atoms)
of the solution. For the case of 3D elastic collisions and the
hard sphere interaction these parts were calculated on the basis
of Eqs.~ (\ref{ad4})-(\ref{ad5}). For the velocities of cells
essentially less, than the characteristic velocity of atoms we can
ignore the velocity-dependent multipliers and consider the parts
of the coefficients $\beta_{el}(V)=\beta_0$ and $D_{el}(V)=D_0$,
connected with the elastic collisions, as the constants. These
constants, as well as the initial velocity-dependent functions
(for $V < v_{Ta}$) (see Eq.(\ref{E})) are connected by the
Einstein relation $D_{el}(V)=M^{-1}T\beta_{el}(V)\simeq
M^{-1}T\beta_0$.

If we make the natural assumption, that $\Sigma(P)$ is a constant,
what means that the PT $w_\varepsilon$ is not dependent on the
cell velocity (as the process, which is determined by the inner
state of the grains), we find from the Eq. (\ref{AC3}), that
$\beta_\varepsilon (P)\sim 1/P$. Due to this specific dependence
the Fokker-Planck equation for cells can be written as

\begin{equation}
\frac{df_g(V)}{dt}= \frac{\partial}{\partial V}\left
\{[V\beta_0-K_\varepsilon]f_g(V)+
[D_0+D_\varepsilon]\frac{\partial f_g(V)}{\partial V}\right\} ,
\label{AC5}
\end{equation}
where $K_\varepsilon>0$ is a constant, determined by the equality
$\beta_\varepsilon\equiv -\frac{K_\varepsilon}{V}$ and the
Eq.(\ref{AC3}), and $D_\varepsilon$ is determined by the
Eq.(\ref{AC4}). For $\beta_0$ can be used the Eq.~(\ref{ad4}).
Finally, the result is $\beta_0 = 8 \sqrt {2T}\, mn_aS /(3M
\sqrt{\pi m})$, where $S=\pi a^2$ is the area of the grain and
$n_a$ is the density of the atoms. The stationary solution of the
Eq.(\ref{AC5}) is the Gaussian distribution:

\begin{equation}
f_g(V)= C exp \left\{ - \frac{\beta_0}{2D_\Sigma}\left(V -
\frac{K_\varepsilon}{\beta_0}\right)^2 \right\}  , \label{AC7}
\end{equation}
Here $C$ is the constant of normalization and $D_\Sigma\equiv
(D_0+D_\varepsilon)$.

The velocity dependence of the distribution function $f_g(V)$
(\ref{AC7}) coincides with one, which has been found in [32] on
basis of the phenomenological assumption, concerning the structure
of the friction coefficient in the Langevin equation. This type of
the velocity distribution function in our consideration is the
consequence of physically clear choice of the perturbation
transition function. It can be generalized for more complicated
and practically important cases, when there are one or more
(additional to ${\bf P}$) vectors, which determine the direction
of ${\bf q}$. It can be some inner vector - "driver", which can be
orientated on the external (e.g., surface) gradients of density,
or temperature, or concentration of some ingredient in the ambient
medium. In that case, naturally, the equilibrium state is not
effectively one-dimensional. Active particle (e.g., cell) can turn
during motion. These problems will be considered separately.

\section{Conclusions}

Here we use the simple and effective way for concretization of the
Fokker-Planck equation on basis of self-consistent determination
of the friction and diffusion coefficients. Both are determined as
the functionals of probability transition. This function possesses
a very different structure for the Boltzmann-type collisions and
the other ones. We found PT for the Boltzmann-type collisions and
proved that velocity dependent friction and diffusion coefficients
are connected by the Einstein relation for the velocities of grain
less than thermal velocity of the small particles. At the same
time there is crucial violation of the Einstein relation for the
higher grain velocity. Therefore in general for the
velocity-dependable friction and diffusion coefficients even for
the Boltzmann-type collisions the applicability of the Einstein
relation is limited by not very high (nevertheless practically
most important) values of the grain velocity. The velocity
dependence of these coefficients and renormalization of the
friction coefficient in 2D and 3D cases, as a consequence of the
tensorial structure of diffusion are found. Because the
Fokker-Planck equation is single-valued also for the
velocity-dependent coefficients the problem of connection between
Langevin and Fokker-Planck equation has to be reformulated as a
problem of the relevant (to the Fokker-Planck equation) Langevin
equation.

For the non-Boltzmann collisions, e.g. for the absorption
collisions, the structure of PT follows from the structure of the
collision integral, obtained earlier [11,14,15] and leads, in
particular, to the different from the Einstein's relations between
the coefficients in the relevant Fokker-Planck equation already
for the region of low grain velocities.

As the example of application of the PT method to the more
complicated systems we considered the generalized friction force
in dusty plasma. The scattering and collecting parts of this force
are determined by the generalized friction coefficient, as a
function of ion stream and grain velocities. The remarkable fact
is that the sign of the collecting friction coefficient can be
negative for some plasma parameters, as it was recently shown
[13]. Of cause realization of negative total friction coefficient
for grains for dusty plasmas in experiment requires the special
conditions, because other mechanisms of friction are exist. The
ion drag force in such approach, as well as the friction force
itself, are the particular cases of this generalized friction. The
ion scattering and collecting drag forces are found and calculated
for the various particular cases. Some phenomenological
expressions, which have been used for calculations before, are
rigorously proved and generalized.

We also constructed the PT for the active particles (e.g., grains
or cells) in an ambient medium for some simple situation. On basis
of physically clear assumptions we found that the part of the
generalized friction coefficient, responsible for self-motion, can
possess the peculiarity $1 \over P $, where P is the momentum of a
grain. Given appropriate also the usual friction mechanism the
stationary solution of the relevant Fokker-Planck equation is
gaussian with a peak around some non-zero velocity. Some
generalizations of the obtained results for more complicated cases
are suggested.

\section*{Acknowledgment}
I am very indebted to Sergei Filimonow for his kind and permanent
support that provided possibility to write this paper.\\
Author thanks E.A.\,Allahyarov, W. Ebeling, U.\, Erdmann,
M.V.\,Fedorov, V.E.\,Fortov, A.M.\,Ignatov, L.\,Schimansky-Geier,
P.P.J.M.Schram, I.M. Sokolov, A.G.\,Zagorodny for valuable
discussions of the various problems, reflected in this work.\\
Author is grateful to G.\,Morfill for invitation in the Max Planck
Institute for Extraterrestrial Physics in Garching and to him and
S.A.\,Khrapak for useful discussion of drag force in dusty
plasmas.

\end{document}